# Gaussian laser beam transformation into an optical vortex beam by helical lens

Ljiljana Janicijevic and Suzana Topuzoski

*Institute of physics, Faculty of natural sciences and mathematics, University "Ss Cyril and Methodius", 1000 Skopje, Republic of Macedonia*

*Corresponding author: suzana_topuzoski@yahoo.com

**Abstract**

In this article we investigate the Fresnel diffraction characteristics of the hybrid optical element which is a combination of a spiral phase plate (SPP) with topological charge $p$ and a thin lens with focal length $f$, named the helical lens (HL). As incident a Gaussian laser beam is treated, having its waist a distance $\zeta$ from the HL plane and its axis passing through the centre of the HL. It is shown that the SPP introduces a phase singularity of $p$-th order to the incident beam, while the lens transforms the beam characteristic parameters. The output light beam is analyzed in detail: its characteristic parameters and focusing properties, amplitude and intensity distributions and the vortex rings profiles and radii, at any $z$ distance behind the HL plane, as well as in the near and far field.

## 1. Introduction

The optical elements with embedded phase patterns in their phase transmission functions have attracted an increased interest because they produce optical vortex beams [1, 2]. The optical vortex beams or optical vortices possess screw dislocations on their wavefronts, where the phase is nondefined, and accordingly, the wave amplitude and intensity are equal to zero. Due to its helicoidal phase wavefront around the singular axis, the optical vortex beam carries orbital angular momentum (OAM) along its propagation axis [3]. The OAM was shown to be proportional to the topological charge (TC) number of the beam. The modulus of the TC shows how many times the phase changes from 0 to $2\pi$ radians in azimuthal direction around the singular axis, while its sign defines the direction of the wavefront helicity.

It was shown that this OAM can be transferred to a captured microparticle causing its rotation in a direction determined by the sign of the TC [4]. The optical vortex beams have been generated by a variety of techniques including a laser cavity [5,6], spiral phase plate (SPP) [7, 8], computer-generated holograms [9,10,11], a magneto-optic spatial light modulator [12], an x-ray angular structure [13] and liquid-crystal displays LCDs [14]. The SPP is an optical element whose optical thickness increases in azimuthal direction, around the plate center, thus, having a transmission function $T(\varphi) = \exp(ip\varphi)$ ($\varphi$ is the azimuthal coordinate and the parameter $p$ governs the total phase shift as the angle changes from 0 to $2\pi$ rad). It is a well known element which transforms the Gaussian beam into output vortex beam [15].

There are other numerous applications of vortex beams; They have been used for optical trapping of particles [16, 17], atom trapping and guiding [18], as information carriers [19] for multiplexing in free-space communications [20], and for realizing of electron vortex beams [21], just to mention a few.

The phase mask with transmission function of the SPP is a useful optical device having applications ranging from astronomy (optical vortex coronagraph) to fiber optics, high power fiber lasers, microscopy, lithography, optical tweezers and quantum optics. [22, 23].

The authors in [24] created vortex-producing lenses by multiplying the angular phase pattern of the SPP with the phase of a Fresnel lens. They experimentally encoded their transmission functions onto a parallel-aligned nematic LCD, and used them in experimental test for optical processing [24].

In [25] the phase function of the vortex lens with singularities embedded at off-axis locations was translated to a spatial light modulator, and the intensity distributions near the focal plane were experimentally obtained.

In this article we study in detail the hybrid optical element-a combination of a SPP with topological charge $p$ and a thin lens with focal length $f$, named the helical lens (HL), in the process of Fresnel diffraction of a Gaussian laser beam. The beam waist of radius $w_0$ is a distance $z = \zeta$ from the HL plane. The analytical theory we present gives results for the amplitude and intensity distributions of the diffracted beam, its characteristic parameters, and the near- and far-field approximations. Further, these results have been specialized for the cases when the SPP is absent (i.e. the Gaussian beam is transferred through the lens), the lens is absent (i.e. the Gassian beam diffracts by the SPP only), and a special geometry when the Gaussian beam enters with its waist in the HL plane.

## 2. The transmission function of the helical lens and the incident beam

The hybrid optical element which is a combination of a spiral phase plate with topological charge $p$ and a thin lens with focal length $f$, named the helical lens, introduces phase retardation to the incident light

$$T(r,\varphi) = \exp\left[-i\left(p\varphi - kr^2/2f\right)\right]. \quad (1)$$

In Fig. 1 the equiphase lines of the HL with topological charge $p=4$ are shown.

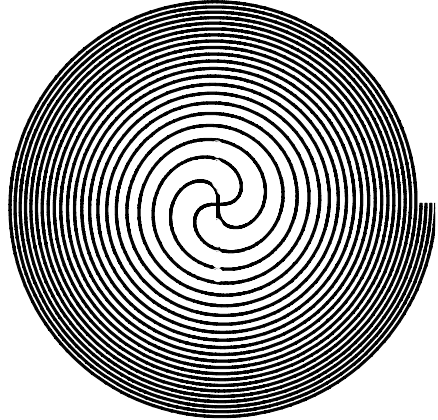

Fig.1. Equiphase lines of the helical lens with TC $p=4$.

This diffractive optical element is illuminated with a Gaussian laser beam, whose waist of radius $w_0$ is a distance $z = \zeta$ from the HL plane. In this plane the light beam amplitude is given by the expression

$$U^{(i)}(r,\varphi,\zeta) = \frac{q(0)}{q(\zeta)} \exp\left\{-ik\left[\zeta + \frac{r^2}{2q(\zeta)}\right]\right\}$$
$$= \frac{w_0}{w(\zeta)} \exp\left(\frac{-r^2}{w^2(\zeta)}\right) \exp\left\{-ik\left[\zeta + \frac{r^2}{2R(\zeta)} + \Psi_0(\zeta)\right]\right\} \qquad (2)$$

where $\Psi_0(\zeta) = (1/k)\arctan(\zeta/\zeta_0)$.

In the previous equation with $k = 2\pi/\lambda$ the wave number is denoted, while $q(\zeta) = \zeta + ikw_0^2/2$ is the beam complex parameter and $\zeta_0 = kw_0^2/2$ is the Rayleigh distance. To the beam's complex curvature $\frac{1}{q(\zeta)} = \frac{1}{R(\zeta)} - \frac{2i}{kw^2(\zeta)}$, a real on-axial radius of curvature $R(\zeta) = \zeta\left[1 + (\zeta_0/\zeta)^2\right]$ is assigned, while $w(\zeta) = w_0\left[1 + (\zeta/\zeta_0)^2\right]^{1/2}$ is the beam transverse amplitude profile radius.

## 3. The theory of Fresnel diffraction of a Gaussian laser beam by HL

In order to calculate analytically the diffracted wave field behind the helical lens, we use the Fresnel Kirchoff diffraction integral [26]

$$U(\rho,\theta,z) = \frac{ik}{2\pi(z-\zeta)} \exp\left\{-ik\left[z - \zeta + \frac{\rho^2}{2(z-\zeta)}\right]\right\}$$
$$\times \int_0^\infty \int_0^{2\pi} T(r,\varphi) U^{(i)}(r,\varphi,\zeta) \exp\left[-i\frac{k}{2}\left(\frac{r^2}{z-\zeta} - \frac{2r\rho\cos(\varphi-\theta)}{z-\zeta}\right)\right] r\,dr\,d\varphi$$

in which the incident beam (2) and HL transmission function (1) are involved. Considering the identity [27]

$$\exp\left[\frac{ik}{z-\zeta} r\rho\cos(\varphi-\theta)\right] = \sum_{s=-\infty}^{\infty} i^s J_s\left(\frac{kr\rho}{z-\zeta}\right) \exp[is(\varphi-\theta)],$$

and doing the integration over the azimuthal coordinate, the expression for the diffracted wave field amplitude can be written in the form

$$U(\rho,\theta,z) = \frac{ik}{(z-\zeta)} \frac{q(0)}{q(\zeta)} \exp\left[-ik\left(z + \frac{\rho^2}{2(z-\zeta)}\right)\right] \exp\left[-ip\left(\theta + \frac{\pi}{2}\right)\right] Y(\rho,\theta). \qquad (3)$$

In Eq. (3) with $Y(\rho,\theta)$ the integral over the radial variable is denoted

$$Y(\rho,\theta) = \int_0^\infty J_p\left(\frac{k\rho}{z-\zeta} r\right) \exp\left\{-\frac{ik}{2}\left[\frac{1}{z-\zeta} - \frac{1}{f} + \frac{1}{q(\zeta)}\right] r^2\right\} r\,dr. \qquad (4)$$

It is an integral of type $Y_{\mu,\nu} = \int_0^\infty J_\mu(br) \exp(-a^2 r^2) r^{\nu-1} dr$, whose solution, when $\text{Re}(2+\mu) > 0$ and $\text{Re}(a^2) > 0$, is

$$Y_{\mu,\nu} = \frac{1}{2a^2} \sqrt{\frac{\pi}{2} \frac{b^2}{8a^2}} \exp\left(-\frac{b^2}{8a^2}\right) \left[I_{(\mu-1)/2}\left(\frac{b^2}{8a^2}\right) - I_{(\mu+1)/2}\left(\frac{b^2}{8a^2}\right)\right]. \qquad (4a)$$

In our case the indices are: $\mu = p$ and $\nu = 2$. While, the parameters values are: $b = \dfrac{k\rho}{z-\zeta}$ and

$$a^2 = \frac{ik}{2}\left[\frac{1}{z-\zeta} - \frac{1}{f} + \frac{1}{q(\zeta)}\right] \text{ or } \frac{1}{2a^2} = \frac{(z-\zeta)q(\zeta)}{ik\{(z-\zeta)[1-q(\zeta)/f] + q(\zeta)\}}.$$

The quotient $b^2/8a^2$ can be written in the following way

$$\frac{b^2}{8a^2} = \frac{ik}{2}\left[\frac{1}{2\left[z-\zeta + \dfrac{q(\zeta)}{1-q(\zeta)/f}\right]} - \frac{1}{2(z-\zeta)}\right]\rho^2 = \frac{ik}{2}\left[\frac{1}{2Q(z-Z_1)} - \frac{1}{2(z-\zeta)}\right]\rho^2, \quad (5)$$

where we have introduced a new complex beam parameter

$$Q(z-Z_1) = z - \zeta + \frac{q(\zeta)}{1-q(\zeta)/f} = z - \zeta + \frac{\zeta + i\zeta_0}{1-(\zeta + i\zeta_0)/f}. \quad (6)$$

Futher, by separating the real and imaginary parts in the previous equation

$$Q(z-Z_1) = z - \zeta - \frac{(\zeta_0^2/f) - \zeta(1-\zeta/f)}{(1-\zeta/f)^2 + (\zeta_0/f)^2} + i\frac{\zeta_0}{(1-\zeta/f)^2 + (\zeta_0/f)^2}$$

we find that the new complex beam parameter can be written in the form

$$Q(z-Z_1) = z - Z_1 + iZ_0, \quad (7)$$

where we have denoted with $Z_1$ and $Z_0$ these expressions

$$Z_1 = \zeta + \frac{(\zeta_0^2/f) - \zeta(1-\zeta/f)}{(1-\zeta/f)^2 + (\zeta_0/f)^2}, \quad (8)$$

$$Z_0 = \frac{\zeta_0}{(1-\zeta/f)^2 + (\zeta_0/f)^2}. \quad (9)$$

To the new beam complex curvature defined as

$$\frac{1}{Q(z-Z_1)} = \frac{1}{R(z-Z_1)} - \frac{2i}{kW^2(z-Z_1)}, \quad (10)$$

a new real on-axis radius of curvature is assigned, which, according to its definition, at distance $z-Z_1$ can be written as

$$R(z-Z_1) = (z-Z_1)\left[1 + \left(\frac{Z_0}{z-Z_1}\right)^2\right]. \quad (11)$$

In Eq. (10) the parameter $W(z-Z_1)$ is analog to the new beam transverse amplitude profile radius at distance $z-Z_1$

$$W(z-Z_1) = W_0\left[1 + \left(\frac{z-Z_1}{Z_0}\right)^2\right]^{1/2}. \quad (12)$$

Its minimum value

$$W_0 = \frac{w_0}{\sqrt{(1-\zeta/f)^2 + (\zeta_0/f)^2}} \quad (13)$$

is found from Eq. (9) when taking into consideration that the new parameter $Z_0$ is defined as $Z_0 = kW_0^2/2$.

Returning to the argument $b^2/8a^2$, given by Eq. (5), and using the results from (6) to (13), we can also write it in the form

$$\frac{b^2}{8a^2} = \frac{ik}{2}\frac{\rho^2}{Q'(z)}. \tag{14}$$

We put the previous Eq. (14), into Eq. (4a) in the Bessel function argument and in the exponential term. While, in the multiplier $\sqrt{(\pi/2)(b^2/8a^2)}$ we will also implement this expression: $\frac{b^2}{8a^2} = \frac{-ik}{4}\frac{\rho^2}{(z-\zeta)}\frac{Q(\zeta-Z_1)}{Q(z-Z_1)}$. By making equivalence between Eq. (5) and Eq. (14) one can find that

$$\frac{1}{Q'(z)} = \frac{1}{2Q(z-Z_1)} - \frac{1}{2(z-\zeta)}. \tag{15}$$

Further, from the general definition of the beam complex curvature we can write

$$\frac{1}{Q'(z)} = \frac{1}{R'(z)} - \frac{2i}{kW'^2(z)}. \tag{16}$$

Equalizing the previous equation (16) with Eq. (15) into which the expression (10) has been involved, leads to the following conclusions

$$\frac{1}{R'(z)} = \frac{1}{2R(z-Z_1)} - \frac{1}{2(z-\zeta)} \tag{17a}$$

and

$$W'^2(z) = 2W^2(z-Z_1). \tag{17b}$$

Finally, the solution of the radial integral (4) can be written as

$$Y(\rho,\theta) = \frac{2i(z-\zeta)^2}{kQ'(z)}\sqrt{\frac{\pi}{2}\frac{ik}{2Q'(z)}}\,\rho\exp\left[-\frac{ik}{2Q'(z)}\rho^2\right]\left[I_{(p-1)/2}\left(\frac{ik}{2}\frac{\rho^2}{Q'(z)}\right) - I_{(p+1)/2}\left(\frac{ik}{2}\frac{\rho^2}{Q'(z)}\right)\right], \tag{18}$$

or, considering Eq. (16) it can also be presented in the following form

$$Y(\rho,\theta) = \frac{(z-\zeta)}{ik}\left[\frac{Q(\zeta-Z_1)}{Q(z-Z_1)}\right]^{3/2}\sqrt{\frac{\pi}{2}\left(\frac{-ik}{4(z-\zeta)}\right)}\,\rho\exp\left[-\frac{ik}{4}\left(\frac{1}{Q(z-Z_1)} - \frac{1}{z-\zeta}\right)\rho^2\right]$$
$$\times\left[I_{(p-1)/2}\left(\frac{ik}{2}\left(\frac{1}{R'(z)} - \frac{2i}{kW'^2(z)}\right)\rho^2\right) - I_{(p+1)/2}\left(\frac{ik}{2}\left(\frac{1}{R'(z)} - \frac{2i}{kW'^2(z)}\right)\rho^2\right)\right] \tag{19}$$

By inserting Eq. (18) in Eq. (3) we get the following form for the outgoing, transformed by the helical lens beam

$$U(\rho,\theta,z) = \frac{q(0)}{q(\zeta)}\frac{(z-\zeta)}{Q'(z)}\sqrt{\frac{ik\pi}{4Q'(z)}}\exp\left[-ik\left(z + \frac{\rho^2}{2(z-\zeta)}\right)\right]\exp[-ip(\theta+\pi/2)]$$
$$\times \rho\exp\left[-\frac{ik}{2Q'(z)}\rho^2\right]\left[I_{(p-1)/2}\left(\frac{ik}{2}\frac{\rho^2}{Q'(z)}\right) - I_{(p+1)/2}\left(\frac{ik}{2}\frac{\rho^2}{Q'(z)}\right)\right] \tag{20}$$

Eq. (20) can also be written in this form (by the use of Eqs. (16) and (17b))

$$U(\rho,\theta,z) = \frac{q(0)}{q(\zeta)}\frac{(z-\zeta)}{Q'(z)}\sqrt{\frac{ik\pi}{4Q'(z)}}\exp\left\{-ik\left[z + \frac{\rho^2}{2(z-\zeta)} + \frac{\rho^2}{2R'(z)}\right]\right\}\exp[-ip(\theta+\pi/2)]$$
$$\times \rho\exp\left[-\frac{\rho^2}{2W^2(z-Z_1)}\right]\left[I_{(p-1)/2}\left(\frac{ik}{2}\frac{\rho^2}{R'(z)} + \frac{\rho^2}{2W^2(z-Z_1)}\right) - I_{(p+1)/2}\left(\frac{ik}{2}\frac{\rho^2}{R'(z)} + \frac{\rho^2}{2W^2(z-Z_1)}\right)\right]$$
$$\tag{21}$$

In the previous equations (20) and (21) we take into consideration that $Q'(z)$ and $R'(z)$ are defined by relations (15), (16), (17a) and (17b), through the parameters $Q(z-Z_1)$, $R(z-Z_1)$, $W(z-Z_1)$, $Z_1$, $Z_0$ and $W_0$, which are the parameters of the beam transformed by the HL. They will be explained and discussed in the next section. The validity of the results (20) and (21) can be checked by their specialization for the cases of lens-only transformation (when $p=0$) and SPP-only transformation (when $1/f \to 0$).

### 3.1. *Specialization of the results when the SPP is absent (p=0)*

Further, we will specialize the expression (20) for the case when the Gaussian beam is transformed only by the lens, i.e. the topological charge value $p=0$. Applying the identity $I_{\pm\frac{1}{2}}(y) = \sqrt{\frac{2}{\pi y}} \begin{cases} \text{sh}y \\ \text{ch}y \end{cases} = \sqrt{\frac{2}{\pi y}} \begin{cases} [\exp(y)-\exp(-y)]/2 \\ [\exp(y)+\exp(-y)]/2 \end{cases}$, we find that Eq. (20) turns into

$$U_L(\rho,\theta,z) = \frac{q(0)}{q(\zeta)} \frac{(z-\zeta)}{Q'(z)} \exp\left[\frac{-\rho^2}{W^2(z-Z_1)}\right] \exp\left[-ik\left(z + \frac{\rho^2}{2R(z-Z_1)}\right)\right] \quad (22)$$

Further, we rearrange the previous equation by making the following substitutions

$$\frac{q(0)}{q(\zeta)} = \frac{w_0}{w(\zeta)} \exp\left(i \arctan\left(\frac{\zeta}{\zeta_0}\right)\right), \quad (22a)$$

$$\frac{1}{Q'(z)} = \frac{-1}{2(z-\zeta)} \frac{Q(\zeta-Z_1)}{Q(z-Z_1)} = \frac{-1}{2(z-\zeta)} \frac{W(\zeta-Z_1)}{W(z-Z_1)} \exp\left[i\left(\arctan\left(\frac{z-Z_1}{Z_0}\right) - \arctan\left(\frac{\zeta-Z_1}{Z_0}\right)\right)\right], (22b)$$

and

$$W(\zeta-Z_1) = W(Z_1)\sqrt{1 + \frac{(\zeta-Z_1)^2}{Z_0^2}} = W_0\sqrt{1 + \frac{(\zeta-Z_1)^2}{Z_0^2}}, \quad (22c)$$

thus, obtaining its final form

$$U_L(\rho,\theta,z) = \sqrt{1 + \left(\frac{\zeta-Z_1}{Z_0}\right)^2} \frac{w_0}{w(\zeta)} \left\{\frac{W_0}{W(z-Z_1)} \exp\left[\frac{-\rho^2}{W^2(z-Z_1)}\right] \exp\left[-ik\left(z + \frac{\rho^2}{2R(z-Z_1)} + \Psi(z)\right)\right]\right\} \quad (23)$$

where $\Psi(z) = \frac{1}{k}\left[-\arctan\left(\frac{\zeta}{\zeta_0}\right) + \arctan\left(\frac{\zeta-Z_1}{z_0}\right) - \arctan\left(\frac{z-Z_1}{z_0}\right)\right]$.

Having in mind that $Z_1$ and $Z_0$ are determined by the expressions (8) and (9), respectively, the multiplier $\sqrt{1+((\zeta-Z_1)/Z_0)^2}(w_0/w(\zeta))$ for given incident beam parameters and focal length of the lens, plays a role of a constant (since it does not contain the variables $\rho$ and $z$). It does not influence the wave distribution in the big brackets, which is recognized as a Gaussian beam, whose waist is at position $z=Z_1$ and its Rayleigh range value is $Z_0$. Further, as it was noted before, the parameter $W(z-Z_1)$, given by Eq. (12), is the diffracted beam transverse amplitude profile radius, while $W_0 = W_{min}(z-Z_1) = W(0)$ is its waist radius (Fig. 2).

The parameter $R(z-Z_1)$, given by Eq. (11), is the phase wavefront radius on the beam propagation axis. At distances $z = Z_1 \pm Z_0$ it reaches minimum value $R(z-Z_1) = R_{min}(z-Z_1) = 2Z_0$; as $z \to Z_1$ (at the beam waist position), $R(z-Z_1) \to \infty$; and, also when $z-Z_1 \gg Z_0$ (far-field approximation), then $R(z-Z_1) \to \infty$.

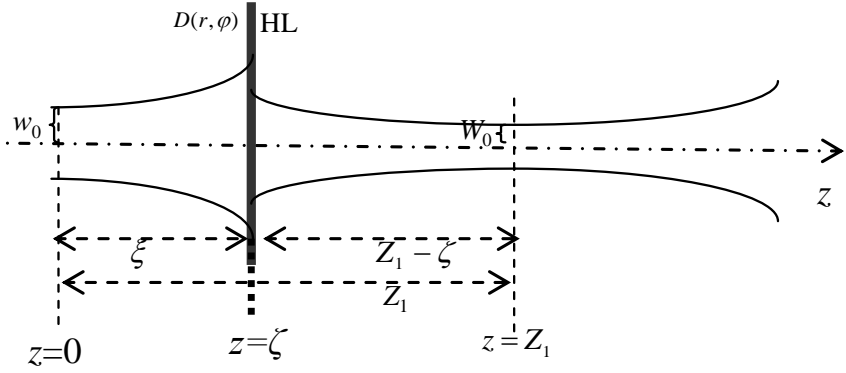

Fig. 2. The characteristic parameters of the incident beam and the diffracted beam.

As it was shown by the relations (8), (9) and (13), the waist position $z = Z_1$, the Rayleigh range value $Z_0$ and the minimum amplitude profile radius $W_0$ of the diffracted beam, depend on the incident beam parameters $\zeta_0, \zeta, w_0$ and $f$. The relations (6)-(13) are actually same with those obtained by Self [28] in his study of a Gaussian beam transformation by a spherical lens of focal length $f$.

Since the lens-transformed beam parameters are at the same time parameters of the beam (21), transformed by the helical lens, we shall briefly discuss their relations. The expression (8) can also be written as

$$\frac{Z_1 - \zeta}{f} = 1 + \frac{\zeta/f - 1}{(\zeta/f - 1)^2 + (\zeta_0/f)^2} \ . \tag{24}$$

It indicates that, when the waist plane of the incident beam is a distance $\zeta > f$ from the lens, then $(Z_1 - \zeta)/f > 1$ i.e. $Z_1 - \zeta > f$, meaning that the waist plane of the transformed beam is at distance $Z_1 > 2f$.

When $\zeta = f$, then $(Z_1 - \zeta)/f = 1$ i.e. $Z_1 = 2f$. The waist plane of the transformed beam is a distance $Z_1 - \zeta = f$ from the $D$-plane, where $z = \zeta$ (Fig. 2). In this case, the waist radius is: $W_0 = w_0 f / \zeta_0$ i.e. $W_0 > w_0$ if $f > \zeta_0$; $W_0 < w_0$ if $f < \zeta_0$, and $W_0 = w_0$ when $f = \zeta_0$.

Equation (24) also indicates that when $\zeta < f$, the waist position of the transformed beam is also smaller than the lens focal length $Z_1 - \zeta < f$. The extreme of this case is when $\zeta = 0$ (the incident beam waist coincides with the $D$-plane). Then

$$Z_0 = \frac{\zeta_0}{1 + (\zeta_0/f)^2} \ , \ W_0 = \frac{w_0}{\sqrt{1 + (\zeta_0/f)^2}} \ \text{and} \ Z_1 = \frac{\zeta_0}{f} \left[\frac{\zeta_0}{1 + (\zeta_0/f)^2}\right] = \frac{\zeta_0}{f} Z_0 \ . \tag{24a}$$

It is obvious that $Z_0 < \zeta_0$ and $W_0 < w_0$. If, in addition $f = \zeta_0$, then it follows: $Z_1 = Z_0 = \zeta_0/2$ and $W_0 = w_0/\sqrt{2}$.

## 3.2. Specialization of the results when the lens is absent

Another specialization of the result (20) i.e. of the parameters $Q(z-Z_1)$, $R(z-Z_1)$, $W(z-Z_1)$, $Z_1$, $Z_0$ and $W_0$ which take part in its analytical formulation, is the case when $f \to \infty$ i.e. when the lens is absent. Then the complex beam parameter defined by Eq. (6) becomes
$$Q(z-Z_1) = z - Z_1 + iZ_0 \xrightarrow[f\to\infty]{} z - \zeta + q(\zeta) = z + i\zeta_0, \quad \text{i.e.} \quad Q(z-Z_1) = q(z) \quad \text{and} \quad Q(\zeta - Z_1) = q(\zeta),$$
because $Z_1 = 0$ and $Z_0 = \zeta_0$ (which is evident from equations (8) and (9)). In this case $W_0 = w_0$, $W(z-Z_1) = w(z) = w_0\sqrt{1+(z/\zeta_0)^2}$, $R(z-Z_1) = R(z) = z[1+(\zeta_0/z)^2]$. All of the parameters appearing in Eq. (20), when the lens is absent, are parameters of the incident Gaussian beam, but defined in the space $z > \zeta$. In that space the beam waist does not occur, but, the presence of the SPP transforms the incident chargeless laser beam into a beam with topological charge $p$, possessing a dark axis. It is represented by the expression

$$U_{SPP}(\rho,\theta,z) = \frac{1}{2}\frac{q(0)}{q(z)}\sqrt{\frac{\pi}{2}\frac{q(\zeta)(-ik)}{(z-\zeta)q(z)}}\exp\left\{-ik\left[z + \frac{\rho^2}{4(z-\zeta)}\right]\right\}\exp[-ip(\theta + \pi/2)]$$
$$\times \rho \exp\left[-\frac{ik}{4q(z)}\rho^2\right]\left[I_{(p-1)/2}\left(\frac{ik}{2}\left(\frac{1}{2q(z)} - \frac{1}{2(z-\zeta)}\right)\rho^2\right) - I_{(p+1)/2}\left(\frac{ik}{2}\left(\frac{1}{2q(z)} - \frac{1}{2(z-\zeta)}\right)\rho^2\right)\right] \quad (25)$$

The solution (25) is characteristic for the vortices produced by the spiral phase plate [15] and for the higher diffraction orders of the forked gratings [29], when they are illuminated by a Gaussian laser beam. It must be outlined that the solution in the reference [15] concerns the case of special incidence, when the waist of the Gaussian incident beam coincides with the diffracting plane i.e. when $\zeta = 0$.

By checking the validity of the expression (20) for the cases when $p=0$ and $(1/f) \to 0$, it became evident that in the transformation process of the Gaussian beam by the helical lens, the role of the lens is to change the incident beam parameters $q(z), w(z), R(z), \zeta_0$ and $w_0$, into a set of new parameters $Q(z-Z_1)$, $W(z-Z_1)$, $R(z-Z_1)$, $Z_0$ and $W_0$, respectively. While, the role of the SPP is to change the chargeless incident Gaussian beam into a vortex beam, now defined by expression (21), possessing the phase singularity of $p$-th order.

## 3.3. Analysis of the field transformed by the HL

The analytical expression of the field (21) written by its real parameters $R(z-Z_1)$ and $W(z-Z_1)$, by using equations (22a) and (22b) is

$$U_{HL}(\rho,\theta,z) = \frac{w_0}{w(\zeta)}\frac{W(\zeta-Z_1)}{W(z-Z_1)}\frac{1}{2}\sqrt{\frac{-ik\pi}{2(z-\zeta)}\frac{W(\zeta-Z_1)}{W(z-Z_1)}}$$
$$\times \exp\left\{-i\left[k\left(z + \frac{1}{4}\left(\frac{1}{R(z-Z_1)} + \frac{1}{z-\zeta}\right)\rho^2\right) + \Phi(z)\right]\right\}\exp[-ip(\theta+\pi/2)] \quad (26)$$
$$\times \rho \exp\left[-\frac{\rho^2}{W'^2(z)}\right]\left[I_{(p-1)/2}\left(\frac{ik}{2}\rho^2\left(\frac{1}{R'(z)} - \frac{2i}{kW'^2(z)}\right)\right) - I_{(p+1)/2}\left(\frac{ik}{2}\rho^2\left(\frac{1}{R'(z)} - \frac{2i}{kW'^2(z)}\right)\right)\right]$$

with $\Phi(z) = -\arctan\left(\frac{\zeta}{\zeta_0}\right) + \frac{3}{2}\left[\arctan\left(\frac{\zeta-Z_1}{z_0}\right) - \arctan\left(\frac{z-Z_1}{z_0}\right)\right]$.

Thus, the intensity distribution, proportional to the squared of the amplitude modulus $I(\rho,z) \propto |U_{HL}(\rho,\theta,z)|^2$ is

$$I(\rho,z) = \frac{k\pi}{8(z-\zeta)} \frac{w_0^2}{w^2(\zeta)} \left[ \frac{W(\zeta - Z_1)}{W(z - Z_1)} \right]^3$$

$$\times \rho^2 \exp\left[ -\frac{2\rho^2}{W'^2(z)} \right] \left| I_{(p-1)/2}\left( \frac{ik}{2}\rho^2 \left( \frac{1}{R'(z)} - \frac{2i}{kW'^2(z)} \right) \right) - I_{(p+1)/2}\left( \frac{ik}{2}\rho^2 \left( \frac{1}{R'(z)} - \frac{2i}{kW'^2(z)} \right) \right) \right|^2 \quad (27)$$

or

$$I_{HL}(\rho,z) = \frac{k\pi}{8(z-\zeta)} \frac{w_0^2}{w^2(\zeta)} \left[ \frac{W(\zeta - Z_1)}{W(z - Z_1)} \right]^3$$

$$\times \rho^2 \exp\left[ -\frac{2\rho^2}{W'^2(z)} \right] \left| J_{(p-1)/2}\left( \frac{k}{2}\rho^2 \left( \frac{1}{R'(z)} - \frac{2i}{kW'^2(z)} \right) \right) - i J_{(p+1)/2}\left( \frac{k}{2}\rho^2 \left( \frac{1}{R'(z)} - \frac{2i}{kW'^2(z)} \right) \right) \right|^2 \quad (28)$$

In the forthcoming Figures 3,4,5 and 6 the radial intensity distribution of the diffracted wave field at different $z$-distances behind the helical lens, calculated by using Eq. (27), and for different values of the incident beam waist position ($\zeta = 250\,\text{mm}$, $\zeta = 500\,\text{mm}$, $\zeta = 750\,\text{mm}$ and $\zeta = 0$, respectively) are presented. For all of them we used the following parameters values: $w_0 = 1\,\text{mm}$, $\lambda = 630\,\text{nm}$, $p = 4$, $f = 500\,\text{mm}$ and $z_0 = 5000\,\text{mm}$. The common conclusion that can be derived is that, the output beam has its waist at distances $Z_1$ defined by Eq. (8) (which have different values for all these four cases). At these axial positions, the intensity drops to zeroth values from the outer sides of the vortex very smoothly; instead, before and after the distance $Z_1$, the sidelobes in the intensity distributions are present. The vortex radius is with smallest value at the waist position, and the intensity 'walls' are sharpest there.

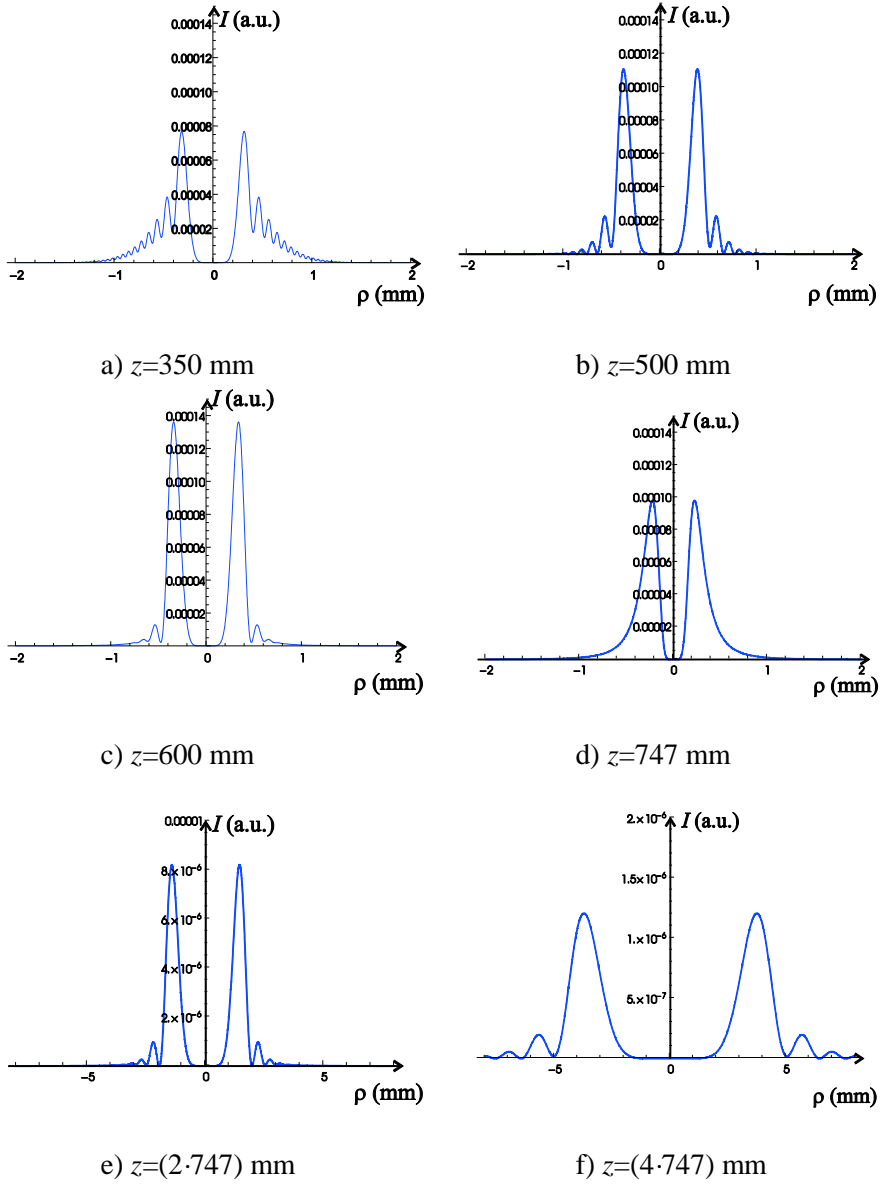

a) $z$=350 mm  b) $z$=500 mm

c) $z$=600 mm  d) $z$=747 mm

e) $z$=(2·747) mm  f) $z$=(4·747) mm

Fig. 3. Radial intensity distribution of the diffracted field at different $z$-distances behind the HL for $w_0 = 1\,\text{mm}$, $\lambda = 630\,\text{nm}$ $p = 4$, $f = 500\,\text{mm}$, $z_0 = 5000\,\text{mm}$ and $\zeta = 250\,\text{mm}$. The beam waist is at $Z_1 = 747\,\text{mm}$.

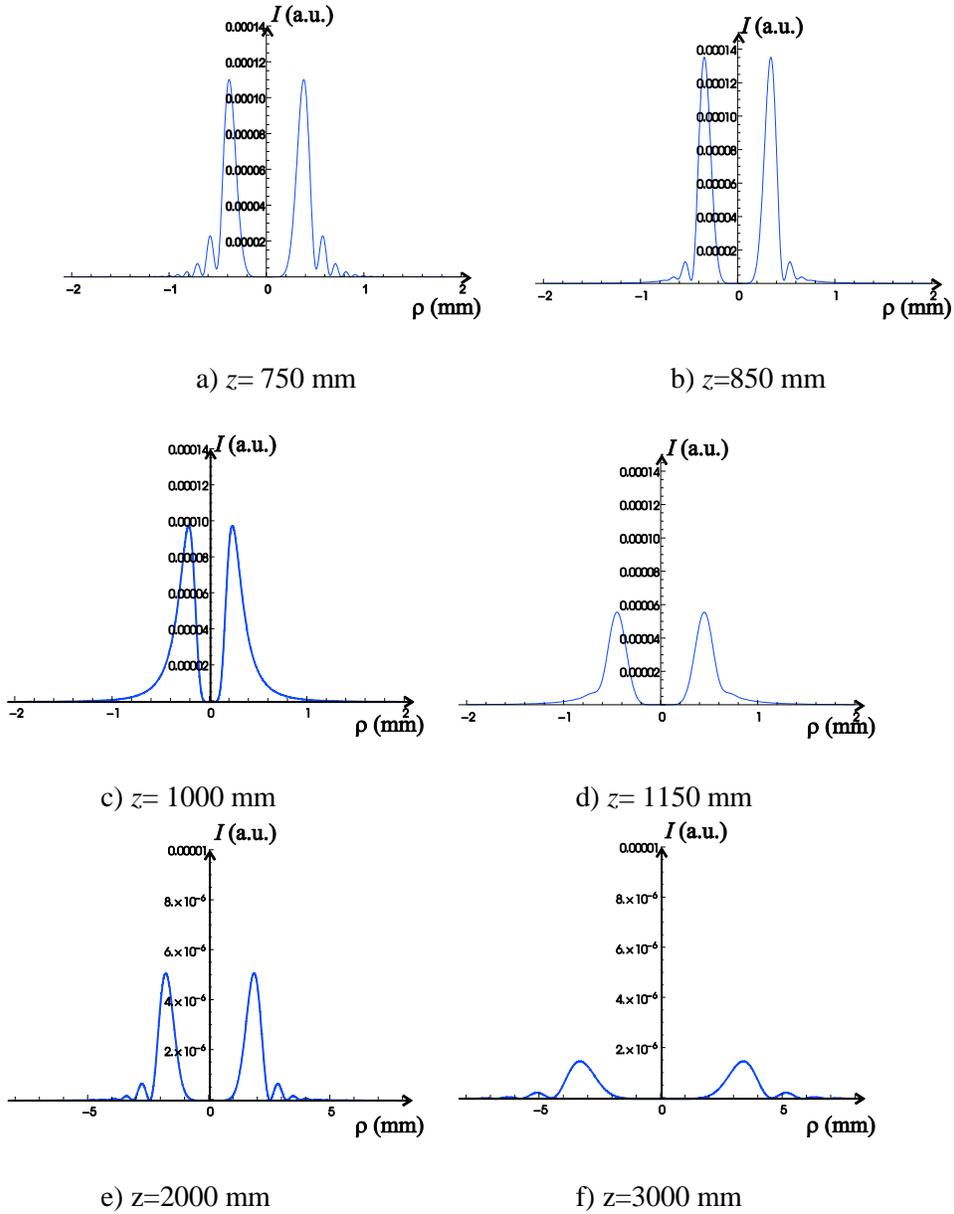

a) $z= 750$ mm  b) $z=850$ mm

c) $z= 1000$ mm  d) $z= 1150$ mm

e) $z=2000$ mm  f) $z=3000$ mm

Fig. 4. Radial intensity distribution of the diffracted field at different $z$-distances behind the HL for $w_0 =1\,\mathrm{mm}$, $p = 4$, $f = 500\,\mathrm{mm}$, $z_0 = 5000\,\mathrm{mm}$ and $\zeta = 500\,\mathrm{mm}$. The beam waist is at $Z_1 = 1000\,\mathrm{mm}$.

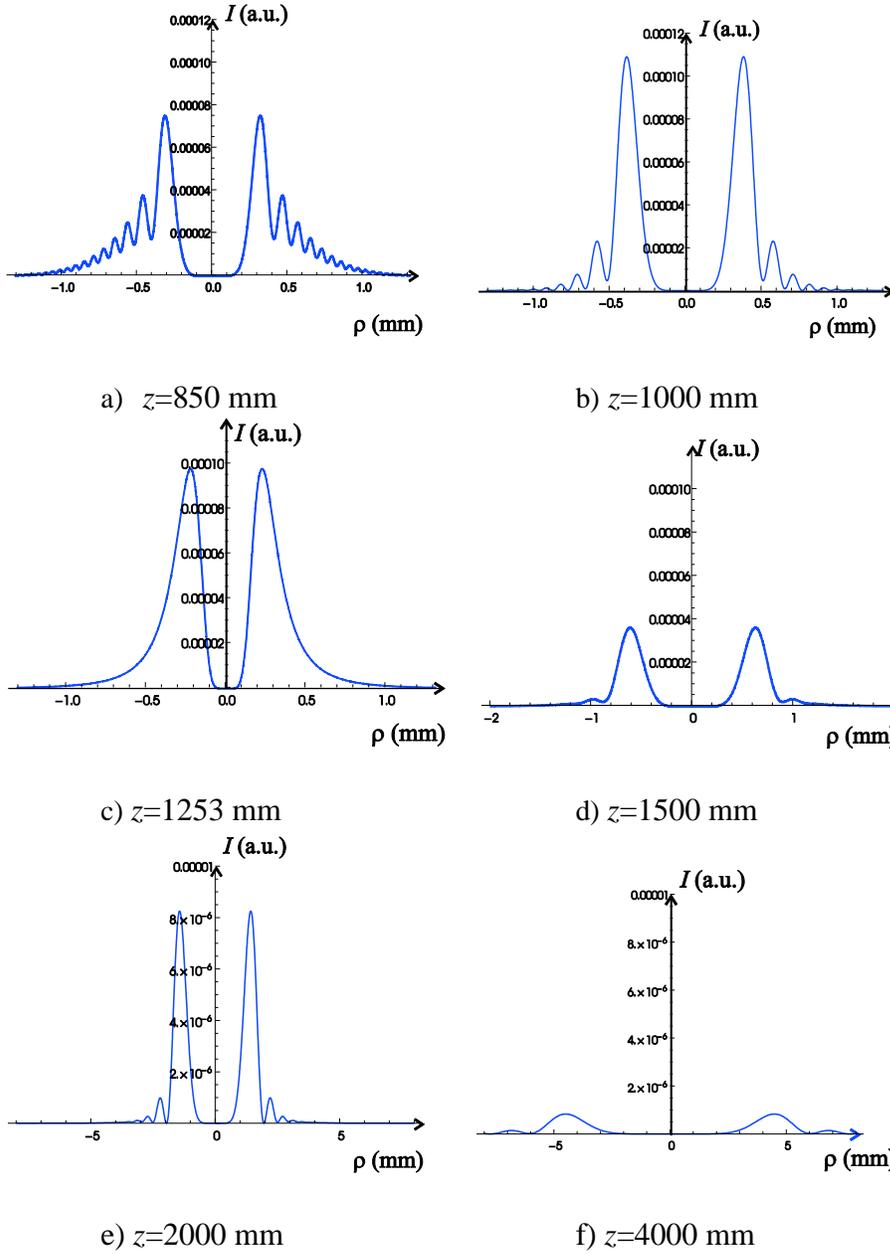

a) $z$=850 mm
b) $z$=1000 mm
c) $z$=1253 mm
d) $z$=1500 mm
e) $z$=2000 mm
f) $z$=4000 mm

Fig. 5. Radial intensity distribution of the diffracted field at different $z$-distances behind the HL for $w_0 = 1\,\text{mm}$, $\lambda = 630\,\text{nm}$ $p = 4$, $f = 500\,\text{mm}$, $z_0 = 5000\,\text{mm}$ and $\zeta = 750\,\text{mm}$. The beam waist is at $Z_1 = 1253\,\text{mm}$.

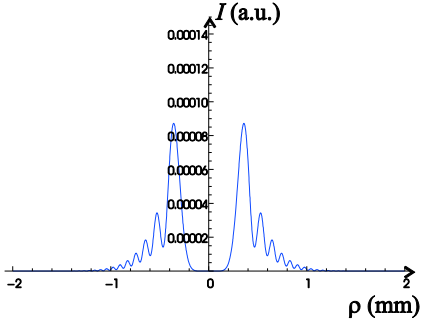
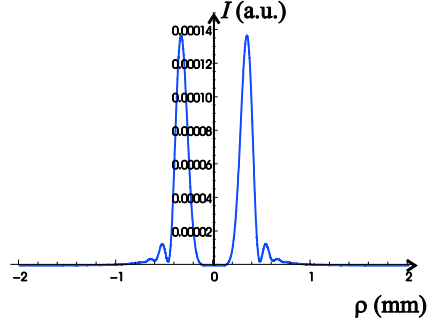

a) $z$=150 mm  b) $z$=350 mm

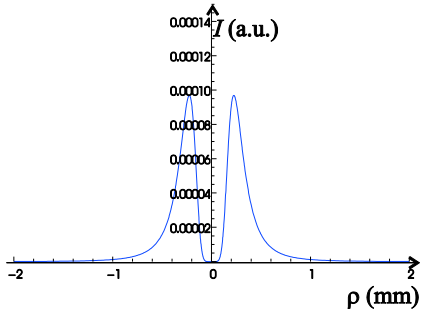
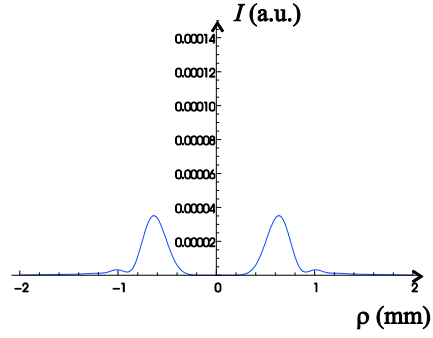

c) $z$=500 mm  d) $z$=750 mm

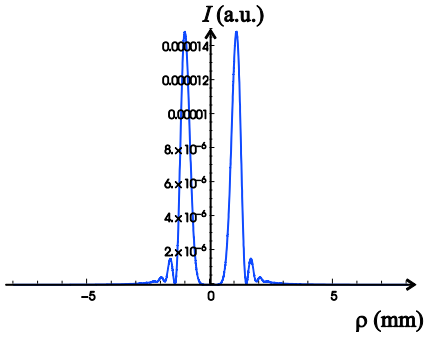
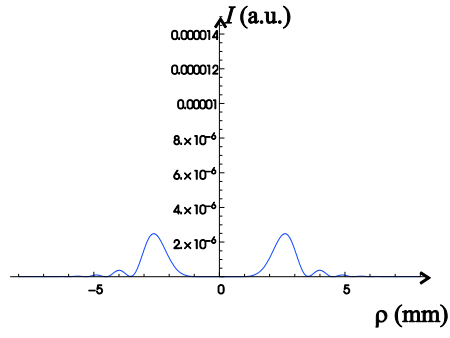

e) $z$=1000 mm  f) $z$=2000 mm

Fig. 6. Radial intensity distribution of the diffracted field at different $z$-distances behind the HL for $w_0 = 1$ mm, $p = 4$, $f = 500$ mm, $z_0 = 5000$ mm and $\zeta = 0$. The beam waist is at $Z_1 = 500$ mm.

In both representations of the diffracted wave field intensity, (27) and (28), the argument of the Bessel functions remains complex. To separate the real and imaginary parts of the difference between two Bessel functions one can employ the Newmann's addition theorem [27] known from the theory of Bessel functions [the separation is shown in the Apendix]. However, in order to find the correct determination of the vortex radius, numerical method for solving the transcendental equation needs to be applied.

Some general information about the spatial behavior of the vortex beams, generated by the helical lens, and the analytical expressions for the vortex radii, can be obtained for very near and far field.

The graphical representation of the functions $1/R(z-Z_1)$, $-1/(z-\zeta)$ and $2/kW^2(z-Z_1)$ helps in forming an opinion about the choice of the regions where entire or approximate presentation of the field $U_{HL}(\rho,z)$ should be taken. To do so, we must have in mind that

$$\frac{2}{kW^2(z-Z_1)} = \frac{1}{Z_0\left[1+((z-Z_1)/Z_0)^2\right]}, \qquad (29)$$

and

$$\frac{1}{R(z-Z_1)} = \frac{1}{(z-Z_1)\left[1+(Z_0/(z-Z_1))^2\right]}, \qquad (30)$$

where the parameters $Z_1$, $Z_0$ and $W_0$ are defined by equations (8), (9) and (13), respectively.

We will choose a special incidence of the Gaussian beam when its waist plane (z=0) is a distance $z=\zeta=f$ from the D-plane (Fig. 2). Then $Z_1=2f$, $Z_0=f^2/\zeta_0$, and $W_0=w_0 f/\zeta_0$. (If, additionally, $f>\zeta_0$, then $W_0>w_0$. While, when $f<\zeta_0$ then $W_0<w_0$, and when $f=\zeta_0$ then $W_0=w_0$).

In the case when $\zeta=f$, it is also valid the following

$$\frac{2}{kW_0^2}=\frac{1}{Z_0}=\frac{\zeta_0}{f^2}; \qquad \frac{1}{z-\zeta}=\frac{1}{z-f}; \qquad \frac{2}{kW^2(z-Z_1)}=\frac{\zeta_0}{f^2\left[1+(\zeta_0(z-2f)/f^2)^2\right]} \qquad \text{and}$$

$$\frac{1}{R(z-Z_1)}=\frac{1}{(z-2f)\left[1+(f^2/\zeta_0(z-2f))^2\right]}. \qquad (30a)$$

When, in addition, $f=\zeta_0=\zeta$, then

$$\frac{1}{z-\zeta}=\frac{1}{z-\zeta_0}; \quad \frac{2}{kW^2(z-Z_1)}=\frac{1}{f\left[1+((z-2f)/f)^2\right]}, \text{ and } \frac{1}{R(z-Z_1)}=\frac{1}{(z-2f)\left[1+(f/(z-2f))^2\right]}$$
(30 b)

In Fig. 7 the variation of the parameters given by Eq. (30b) along the z axis is shown.

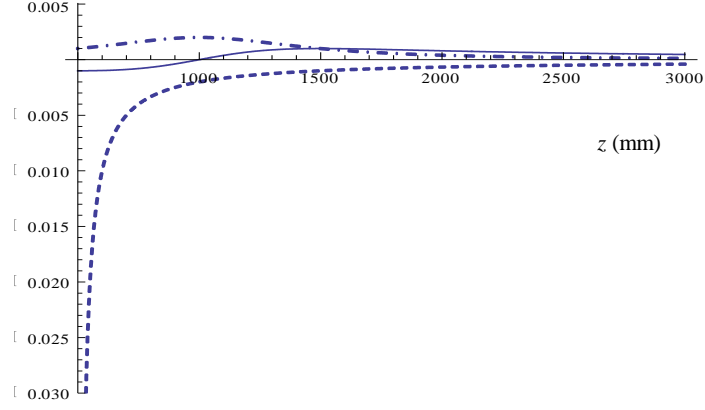

$f = 500\,\text{mm}$

$\zeta = \zeta_0 = f$

$(z-\zeta)^{-1}$ – dashed curve

$(R(z-Z_1))^{-1}$ – solid curve

$2/kW^2(z-Z_1)$ – dot-dashed curve

Fig. 7. The variation of the parameters: $(z-\zeta)^{-1}$ (dashed curve), $(R(z-Z_1))^{-1}$ (solid curve) and $2/kW^2(z-Z_1)$ (dot-dashed curve), given by Eq. (30b), along the $z$ axis. ($f = 500\,\text{mm}$ and $\zeta = \zeta_0 = f$)

**The near field approximation**

From the graph in Fig. 7 it is evident that, in the region close to the diffraction plane the value of the curvature $-1/(z-\zeta)$ dominates over the values of the real $1/R(z-Z_1)$ and imaginary part $2/kW^2(z-Z_1)$ of the complex curvature. Thus, the intensity (28) is approximated as

$$I_{HL}(\rho,z) \approx \frac{k\pi}{8(z-\zeta)} \frac{w_0^2}{w^2(\zeta)} \left[\frac{W(\zeta-Z_1)}{W(z-Z_1)}\right]^3 \rho^2 \left[ J_{(p-1)/2}^2\left(\frac{k\rho^2}{4(z-\zeta)}\right) + J_{(p+1)/2}^2\left(\frac{k\rho^2}{4(z-\zeta)}\right) \right]. \qquad (31)$$

Inserting $x = k\rho^2/4(z-\zeta)$, Eq. (31) can also be written as

$$I_{HL}(\rho,z) \approx \frac{\pi}{2} \frac{w_0^2}{w^2(\zeta)} \left[\frac{W(\zeta-Z_1)}{W(z-Z_1)}\right]^3 x\left[ J_{(p-1)/2}^2(x) + J_{(p+1)/2}^2(x) \right]. \qquad (32)$$

Deriving the intensity over the radial variable and after that making it equal to zero, we arrive to the following transcendental equation: $|J_{(p-1)/2}(x)| = |J_{(p+1)/2}(x)|$. (During the previous calculations we used the relation for the Bessel functions $J_{n-1}(x) + J_{n+1}(x) = (2n+1)x^{-1}J_n(x)$). Further, considering that the argument $x$ has big value when the distance $z-\zeta$ tends to zero, the high-argument value approximation for the Bessel functions: $J_\nu(x) \approx \sqrt{2/\pi x}\cos(x - \nu\pi/2 - \pi/4)$ can be applied in the previous transcendental equation, leading to the result: $x = (p+1)\pi/4$, or, for the vortex radius expression we get

$$\rho = \sqrt{(p+1)\lambda(z-\zeta)/2} \qquad (33)$$

Near to the vortex core (when $\rho \to 0$ we use the small-argument approximation for the Bessel function $J_\nu(x) \approx (x/2)^\nu / \Gamma(\nu+1)$ ($x \to 0$; $\nu \neq -1, -2, ...$), and find that the intensity here drops down to zero as $I \approx \rho^{2p}$. (exactly as $I \approx \dfrac{\rho^{2p}}{(z-\zeta)^{2p}}\left[1 + \dfrac{k^2\rho^4}{16(z-\zeta)^2(p-1)^2}\right]$).

Whereas, far from the vortex core the function $x\left[J^2_{(p-1)/2}(x) + J^2_{(p+1)/2}(x)\right]$ tends to one, so, the intensity tends to constant $I_{HL}(\rho, z) \approx \dfrac{\pi}{2} \dfrac{w_0^2}{w^2(\zeta)} \left[\dfrac{W(\zeta - Z_1)}{W(z - Z_1)}\right]^3$ at some given $z$ distance.

### The far-field approximation

In the region $z \gg Z_1 + Z_0$ the curvature $1/R(z - Z_1) \to +0$, while $-1/(z-\zeta) \to -0$. Their sum tends to zero since they cancel each other.

Therefore, the far-field approximation of the expression (27), since $1/R'(z) = 0$, is

$$I_{HL}(\rho, z) = \dfrac{k\pi}{4(z-\zeta)} \dfrac{w_0^2}{w^2(\zeta)} \dfrac{W^3(\zeta - Z_1)}{W(z - Z_1)} \dfrac{\rho^2}{W'^2(z)} \exp\left[-\dfrac{2\rho^2}{W'^2(z)}\right]$$

$$\times \left\{I_{(p-1)/2}\left(\dfrac{\rho^2}{W'^2(z)}\right) - I_{(p+1)/2}\left(\dfrac{\rho^2}{W'^2(z)}\right)\right\}^2 \qquad (34)$$

To find the radius of the light core surrounding the axial dark region, we deal with the variable $x = \rho^2 / W'^2(z) = \rho^2 / 2W^2(z - Z_1)$ and put equal to zero the first derivative of the intensity expression (34). Since we search for the first maximum in the paraxial region, we will use the small value approximation for the modified Bessel functions $I_\nu(x) \approx (x/2)^\nu / \Gamma(\nu+1)$. Also, for the paraxial region $(\rho \to 0)$ we take that $\exp(-\rho^2/W'^2(z)) \to 1$, and from (34) we find that, towards the vortex axis the intensity drops to zero as $\rho^{2p}$: $I_{HL}(\rho \to 0, z) \propto \dfrac{\rho^{2p}}{W'^2(z)} \left[1 - \dfrac{\rho^2}{(p+1)W'^2(z)}\right]$. From this intensity distribution we have calculated its first maximum at radial position

$$\rho_{max} = W'(z)\sqrt{\dfrac{p(p+1)}{p+2}} . \qquad (35)$$

By the previous Eq. (35) the vortex radii in the far field are defined.
Outside the light hyperboloid we use the large argument approximation for the modified Bessel functions in (34), $I_\nu(x) \approx (2\pi x)^{-1/2} \exp(x)\left[1 - (4\nu^2 - 1)/8x\right]$, and find out that the intensity decreases with $\rho^{-4}$: $I_{HL}(\rho \to \infty, z \gg Z_1 + Z_0) \propto \dfrac{kW^3(\zeta - Z_1)W^3(z - Z_1)}{8(z - \zeta)} \dfrac{p^2}{\rho^4}$.

### 3.4. Specialization of the results for the case of incident beam whose waist is in the HL plane ($\zeta = 0$)

In the case when $\zeta = 0$, then the parameters $Z_0$, $W_0$ and $Z_1$ are defined by equations (24a). Substituing their values in Eq.(26) one gets the amplitude of the diffracted by the HL wave as

$$U_{HL}(\rho,\theta,z) = \frac{w_0}{W(z-Z_1)}\sqrt{\frac{-ik\pi}{8z}}\frac{w_0}{W(z-Z_1)}\exp\left\{-i\left[k\left(z+\frac{1}{4}\left(\frac{1}{R(z-Z_1)}+\frac{1}{z}\right)\rho^2\right)\right]\right\}\exp[-ip(\theta+\pi/2)]$$

$$\times \rho\exp\left[-\frac{\rho^2}{2W^2(z-Z_1)}\right]\left[I_{(p-1)/2}\left(\frac{ik}{2}\rho^2\left(\frac{1}{2R(z-Z_1)}-\frac{1}{2z}\right)+\frac{\rho^2}{2W^2(z-Z_1)}\right)\right.$$

$$\left.-I_{(p+1)/2}\left(\frac{ik}{2}\rho^2\left(\frac{1}{2R(z-Z_1)}-\frac{1}{2z}\right)+\frac{\rho^2}{2W^2(z-Z_1)}\right)\right] \tag{36}$$

When additionally the lens is absent, then $W_0 = w_0$, $Z_0 = \zeta_0$, $Z_1 = 0$, $W(z-Z_1) = w(z)$, $R(z-Z_1) = R(z)$, and accordingly, the amplitude of the beam diffracted by the SPP is

$$U_{SPP}(\rho,\theta,z) = \sqrt{\frac{-ik\pi}{8z}}\exp\left\{-i\left[k\left(z+\frac{1}{4}\left(\frac{1}{R(z)}+\frac{1}{z}\right)\rho^2\right)\right]\right\}\exp[-ip(\theta+\pi/2)]\rho\exp\left[-\frac{\rho^2}{2w^2(z)}\right]$$

$$\times\left[I_{(p-1)/2}\left(\frac{ik}{2}\rho^2\left(\frac{1}{2R(z)}-\frac{1}{2z}\right)+\frac{\rho^2}{2w^2(z)}\right)-I_{(p+1)/2}\left(\frac{ik}{2}\rho^2\left(\frac{1}{2R(z)}-\frac{1}{2z}\right)+\frac{\rho^2}{2w^2(z)}\right)\right] \tag{37}$$

and its far-field approximation (valid in the region $z \gg \zeta_0$) is

$$U_{SPP}(\rho,\theta,z) = \sqrt{\frac{-ik\pi}{8z}}\exp\left\{-i\left[k\left(z+\frac{1}{4z}\rho^2\right)\right]\right\}\exp[-ip(\theta+\pi/2)]\rho\exp\left[-\frac{\rho^2}{2w^2(z)}\right]$$

$$\times\left[I_{(p-1)/2}\left(\frac{\rho^2}{2w^2(z)}\right)-I_{(p+1)/2}\left(\frac{\rho^2}{2w^2(z)}\right)\right] \tag{38}$$

By making the previous approximations, for the incident beam waist position $\zeta = 0$ and when the lens is absent i.e. $f \to \infty$, we have obtained the results same as those in [15].

## 4. Conclusions

In this work we have presented the complete theoretical study about the Gaussian laser beam transformation through a helical lens (an optical element which is a combination of a spiral phase plate and a converging lens). A similar optical element, a hybrid between SPP and axicon, named the helical axicon, was proposed and its diffraction characteristics when being illuminated by a Gaussian laser beam were investigated by the authors in [30]. In [31] it was shown that the helical axicon converts the diverging vortex LG beam into nondiverging Bessel beam by changing its TC.

The analytical expressions for the diffracted wave field amplitude and intensity, by the helical lens, are derived in the form of a difference between two modified Bessel functions of complex argument.

By specializing the wave field amplitude, obtained by diffracting the Gaussian beam by helical lens, for the case of absence of the lens, we have shown that, in the transformation process the role of the SPP with TC $p$ is to introduce a phase singularity of the same order to the output beam.

While, by specializing the wave field amplitude for the case when the SPP is absent, one can see that the role of the lens is to change the incident beam parameters $q(z), w(z), R(z), \zeta_0$ and $w_0$, into a set of new parameters $Q(z-Z_1), W(z-Z_1), R(z-Z_1), Z_0$ and $W_0$, respectively.

The output beam, obtained by Fresnel diffraction of a Gaussian beam by HL, has a waist at axial distance $Z_1$ where the vortex radius is with smallest value, as shown in the graphical representations of the analytically obtained results.

Then, we have calculated the near and far field approximations of the intensity distribution, from where we have obtained the analytical formulas for the vortex rings radii.

On the end, the diffracted wave field transformed by the HL was specialized for a specific case of incidence of the Gaussian beam, when its waist is in the HL plane.

The results from this work could find interest and application in many fields where the optical vortices are implemented. For instance, they can be applied in optical trapping setups, where one can arrange the focusing axial position by dealing with incident beam parameters $\zeta$ and $\zeta_0$ or the lens focal length *f*. Also, the desirable radius of the vortex core can be achieved, as well as the needed value of the orbital angular momentum, which can be transferred to the trapped particle.

**Appendix**

To separate the real and imaginary parts of the difference between two modified Bessel functions we will employ the addition theorem known from the theory of Bessel functions. According to reference [32] the identity

$$\sum_{s=-\infty}^{\infty}(\pm 1)^s \cos(s\alpha) I_s(x) I_{s+\nu}(y) = \cos\left(\nu \arcsin \frac{x \sin \alpha}{\sqrt{x^2 + y^2 \pm 2xy \cos \alpha}}\right) I_\nu\left(\sqrt{x^2 + y^2 \pm 2xy \cos \alpha}\right) \quad (I)$$

specialized for $\alpha = 0$, and if the upper sign is considered, gives

$$I_\nu(x+y) = \sum_{s=-\infty}^{\infty} I_s(x) I_{s+\nu}(y) \quad (II)$$

Therefore the expression (27) turns into

$$I_{HL}(\rho, z) = \frac{k\pi}{8(z-\zeta)} \frac{w_0^2}{w^2(\zeta)} \left[\frac{W(\zeta - Z_1)}{W(z - Z_1)}\right]^3 \rho^2 \exp\left[-\frac{2\rho^2}{W'^2(z)}\right]$$

$$\times \left\{ \left[\sum_l (-1)^l J_{2l}\left(\frac{k\rho^2}{2R'(z)}\right)\left(I_{2l+(p-1)/2}\left(\frac{\rho^2}{W'^2(z)}\right) - I_{2l+(p+1)/2}\left(\frac{\rho^2}{W'^2(z)}\right)\right)\right]^2 \right.$$

$$\left. + \left[\sum_l (-1)^l J_{2l+1}\left(\frac{k\rho^2}{2R'(z)}\right)\left(I_{2l+1+(p-1)/2}\left(\frac{\rho^2}{W'^2(z)}\right) - I_{2l+1+(p+1)/2}\left(\frac{\rho^2}{W'^2(z)}\right)\right)\right]^2 \right\} \quad (III)$$

To separate the real and imaginary parts of the Bessel functions $J_{\frac{p\mp 1}{2}}(\ldots)$ in (28) we apply further the Newmann's addition theorem [27]

$J_\nu(x-y) = \sum_{s=-\infty}^{\infty} J_s(x) J_{s+\nu}(y)$, when $|y/x| < 1$, and obtain

$$I_{HL}(\rho, z) = \frac{k\pi}{8(z-\zeta)} \frac{w_0^2}{w^2(\zeta)} \left[\frac{W(\zeta - Z_1)}{W(z - Z_1)}\right]^3 \rho^2 \exp\left[-\frac{2\rho^2}{W'^2(z)}\right]$$

$$\times \left\{ \left[\sum_l (-1)^l \left(I_{2l}\left(\frac{\rho^2}{W'^2(z)}\right) J_{2l+(p-1)/2}\left(\frac{k\rho^2}{2R'(z)}\right) + I_{2l+1}\left(\frac{\rho^2}{W'^2(z)}\right) J_{2l+1+(p+1)/2}\left(\frac{k\rho^2}{2R'(z)}\right)\right)\right]^2 \right.$$

$$\left. + \left[\sum_l (-1)^l \left(I_{2l}\left(\frac{\rho^2}{W'^2(z)}\right) J_{2l+(p-1)/2}\left(\frac{k\rho^2}{2R'(z)}\right) - I_{2l+1}\left(\frac{\rho^2}{W'^2(z)}\right) J_{2l+1+(p+1)/2}\left(\frac{k\rho^2}{2R'(z)}\right)\right)\right]^2 \right\} \quad (IV)$$

The donut-Gaussian term $\rho^2 \exp\left[-\rho^2/W'^2(z)\right]$ enables zero axial intensity surrounded by light core. This distribution is additionally modified by the squared sums of Bessel and modified Bessel functions, whose orders depend on the TC value $p$.

**References**

[1] J. F. Nye and M. V. Berry, "Dislocations in wave trains," Proc. R. Soc. London Ser. A 336 (1974) 165–190.
[2] M. R. Dennis, K. O'Holleran, and M. J. Padgett, Chapter 5 Singular Optics: Optical Vortices and Polarization Singularities, Progress in Optics 53 (2009) 293-363.
[3] L. Allen, M. W. Beijersbergen, R. J. C. Spreeuw, and J. P. Woerdman, "Orbital angular momentum of light and the transformation of Laguerre-Gaussian laser modes", Phys. Rev. A 45 (1992) 8185-8189.
[4] H. He, M. E. J. Friese, N. R. Heckenberg, and H. Rubinsztein-Dunlop, "Direct observation of transfer of angular momentum to absorptive particles from a laser beam with a phase singularity," Phys. Rev. Lett. 75 (1995) 826-829.
[5] P. Coullet, L. Gil, and F. Rocca, "Optical vortices," Opt. Commun. 73 (1989) 403–408.
[6] M. Harris, C. A. Hill, and J. M. Vaughan, "Optical helices and spiral interference fringes," Opt. Commun. 106 (1994) 161–166.
[7] S. N. Khonina, V. V. Kotlyar, M. V. Shinkaryev, V. A. Soifer, and G. V. Uspleniev, "The phase rotor filter", J. Mod. Opt. 39 (1992) 1147-1154.
[8] M. W. Beijersbergen, R. P. C. Coerwinkel, M. Kristensen, and J. P. Woerdman, "Helical-wavefront laser beams produced with a spiral phase plate", Opt. Commun. 112 (1994) 321-327.
[9] V. Yu, Bazhenov, M. V. Yasnetsov, and M. S. Soskin, "Laser beams with screw dislocations in their wavefronts," JETP Lett. 52 (1990) 428–429.
[10] N. R. Heckenberg, R. McDuff, C. P. Smith, and A. G. White, "Generation of optical phase singularities by computer-generated holograms," Opt. Lett. 17 (1992) 221–223.
[11] Z. Jaroszewicz and A. Kolodziejczyk, "Zone plates performing generalized Hankel transforms and their metrological applications," Opt. Commun. 102 (1993) 391–396.
[12] J. A. Davis, E. Carcole, and D. M. Cottrell, "Intensity and phase measurements of nondiffracting beams generated with a magneto-optic spatial light modulator," Appl. Opt. 35 (1996) 593–598.
[13] A. G. Peele, P. J. McMahon, D. Paterson, C. Q. Tran, A. P.Mancuso, K. A. Nugent, J. P. Hayes, E. Harvey, B. Lai, and I. McNulty, "Observation of an x-ray vortex," Opt. Lett. 27 (2002) 1752–1754.
[14] M. Reicherter, T. Haist, E. U. Wageman, and H. J. Tiziani, "Optical particle trapping with computer-generated holograms written on a liquid-crystal display," Opt. Lett. 24 (1999) 608–610.
[15] V. V. Kotlyar, A. A. Almazov, S. N. Khonina, V. A. Sofier, H. Elfstrom, and J. Turunen, "Generation of phase singularity through diffracting a plane or Gaussian beam by a spiral phase plate", J. Opt. Soc. Am. A 22 (2005) 849-861.
[16] H. He, N. R. Heckenberg, and H. Rubinsztein-Dunlop, "Optical particle trapping with higher-order doughnut beams produced using high efficiency computer generated holograms," J. Mod. Opt. 42 (1995) 217–223.
[17] K. T. Gahagan and G. A. Swartzlander, Jr., "Optical vortex trapping of particles," Opt. Lett. 21 (1996) 827–829.
[18] N. Friedman, A. Kaplan, and N. Davidson, "Dark optical traps for cold atoms," in Advances in Atomic, Molecular and Optical Physics (Elsevier Science, 2002), pp. 101–106.
[19] G. Molina-Terriza, J. P. Torres, and L. Torner, "Twisted photons," Nat. Phys. 3 (2007) 305–310.
[20] J. Wang, J.-Yu. Yang, I. M. Fazal, N. Ahmed, Y. Yan, H. Huang, Y. Ren, Y. Yue, S. Dolinar, M. Tur, and A. E. Willner, "Terabit free-space data transmission employing orbital angular momentum multiplexing," Nat. Photonics 6 (2012) 488–496.
[21] R. Van Boxem, J.Verbeeck and B. Partoens, "Spin effects in electron vortex states," Europhysics Letters 102 (2013) 40010.
[22] G. A. Swartzlander, Jr., "The optical vortex lens," Optics &Photonics News (November, 2006) 39-43.
[23] G. A. Swartzlander, Jr., "Achromatic optical vortex lens," Opt. Lett. 31 (2006) 2042-2044.

[24] K. Crabtree, J. A. Davis and I. Moreno, "Optical processing with vortex-producing lenses," Applied Optics 43 (2004) 1360-1367.

[25] S. Vyas, R. Kumar Singh, D. Pal Ghai, and P. Senthilkumaran1, "Fresnel lens with embedded vortices," International Journal of Optics (2012) doi:10.1155/2012/249695.

[26] M. Born and E. Wolf, Principles of Optics (Cambridge U. Press, 1999).

[27] M. Abramowitz and I. A. Stegun, Handbook of mathematical functions (Dover publ. Inc.,1964).

[28] S. A. Self, "Focusing of spherical Gaussian beams," Applied Optics 22 (1983) 658-661.

[29] Lj. Janicijevic and S. Topuzoski, "Fresnel and Fraunhofer diffraction of a Gaussian laser beam by fork-shaped gratings", Journal of the Optical Society of America A 25 (2008) 2659-2669.

[30] V. V. Kotlyar, A. A. Kovalev, R. V. Skidanov, O. Yu. Moiseev, and V. A. Soifer, "Diffraction of a finite-radius plane wave and a Gaussian beam by a helical axicon and a spiral phase plate," J. Opt. Soc. Am. A 24 (2007) 1955-1964.

[31] S. Topuzoski and Lj. Janicijevic, "Conversion of high order Laguerre-Gaussian beams into Bessel beams of increased, reduced or zero-th order by use of a helical axicon," Opt. Commun. 282, (2009) 3426-3432.

[32] A. P. Prudnikov, Y. A. Brichkov and O. I. Marichev, Integrals and Series; Special Functions (Nauka, Moskva, 1983).

**Figure captions:**

Fig.1. Equiphase lines of the helical lens with TC *p*=4.

Fig. 2. The characteristic parameters of the incident beam and the diffracted beam.

Fig. 3. Radial intensity distribution of the diffracted field at different *z*-distances behind the HL for $w_0 = 1\,\text{mm}$, $\lambda = 630\,\text{nm}$ $p = 4$, $f = 500\,\text{mm}$, $z_0 = 5000\,\text{mm}$ and $\zeta = 250\,\text{mm}$. The beam waist is at $Z_1 = 747\,\text{mm}$.

Fig. 4. Radial intensity distribution of the diffracted field at different *z*-distances behind the HL for $w_0 = 1\,\text{mm}$, $p = 4$, $f = 500\,\text{mm}$, $z_0 = 5000\,\text{mm}$ and $\zeta = 500\,\text{mm}$. The beam waist is at $Z_1 = 1000\,\text{mm}$.

Fig. 5. Radial intensity distribution of the diffracted field at different *z*-distances behind the HL for $w_0 = 1\,\text{mm}$, $\lambda = 630\,\text{nm}$ $p = 4$, $f = 500\,\text{mm}$, $z_0 = 5000\,\text{mm}$ and $\zeta = 750\,\text{mm}$. The beam waist is at $Z_1 = 1253\,\text{mm}$.

Fig. 6. Radial intensity distribution of the diffracted field at different *z*-distances behind the HL for $w_0 = 1\,\text{mm}$, $p = 4$, $f = 500\,\text{mm}$, $z_0 = 5000\,\text{mm}$ and $\zeta = 0$. The beam waist is at $Z_1 = 500\,\text{mm}$.

Fig. 7. The variation of the parameters: $(z - \zeta)^{-1}$ (dashed curve), $(R(z - Z_1))^{-1}$ (solid curve) and $2/kW^2(z - Z_1)$ (dot-dashed curve), given by Eq. (30b), along the *z* axis. ($f = 500\,\text{mm}$ and $\zeta = \zeta_0 = f$)